\documentstyle[12pt]{article}
\begin{document}
\hfill{NCKU-HEP-98-11}\par
\hfill{IP-ASTP-10-98}\par
\hfill{hep-ph/9809553}
\vskip 0.5cm
\begin{center}
{\large {\bf Perturbative QCD analysis of $B(B\to Xl\bar{\nu})$, \\
charm yield $\langle n_c \rangle$ in $B$ decay,
and $\tau (\Lambda_b)/\tau (B_d)$}}
\vskip 0.8cm
We-Fu Chang
\vskip 0.1cm
Department of Physics, National Tsing-Hua University,\par
Hsinchu, Taiwan 300, Republic of China
\vskip 0.3cm
Hsiang-nan Li
\vskip 0.1cm
Department of Physics, National Cheng-Kung University,\par
Tainan, Taiwan 701, Republic of China
\vskip 0.3cm
Hoi-Lai Yu \vskip0.1cm
Institute of Physics, Academia Sinica, \par
Taipei, Taiwan 115, Republic of China
\end{center}
\vskip 0.5cm


\baselineskip=2\baselineskip
\vskip 0.5cm
\centerline{\bf Abstract}
\vskip 0.3cm

We apply perturbative QCD factorization theorems to inclusive heavy hadron
decays, and obtain simultaneously a low semileptonic
branching ratio $B(B\to Xl\bar{\nu})=10.16\%$, the
average charm yield $\langle n_c \rangle=1.17$ per $B$ decay,
a small lifetime ratio $\tau(\Lambda_b)/\tau(B_d)=0.78$, and the
correct absolute decay widths of the $B$ meson and of the
$\Lambda_b$ baryon.

\newpage

Dynamics of heavy hadrons is greatly simplified in the limit of
infinite quark mass. Heavy quark effective theory (HQET) provides
a systematic expansion of QCD in this limit, which makes possible
an almost-first-principle analysis of inclusive heavy hadron
decays. Two different QCD-based approaches have been developed to
study these decays: the HQET-based operator product expansion (OPE)
\cite{aa} and perturbative QCD (PQCD) factorization theorems \cite{ab}. In
the OPE formalism the relevant hadronic matrix elements are expanded
in inverse powers of the heavy quark mass $M_Q$. By contrast, in
the PQCD formalism the factorization of a decay rate into a hard
subamplitude and a universal heavy hadron distribution function
is performed according to inverse powers of the heavy hadron mass
$M_H$. It has been shown that to order $1/M_Q$ the two approaches
give identical results in the case of inclusive semileptonic
decays \cite{ab}.

One of the puzzles in heavy hadron decays is the explanation of the low
lifetime ratio $\tau(\Lambda_b)/\tau(B_d)=0.79\pm 0.06$ \cite{ad}. The OPE
prediction to $O(1/M_b^2)$, $M_b$ being the $b$ quark mass, is about 0.99
\cite{NS}. When including the $O(1/M_b^3)$ corrections, the ratio depends
on six unknown parameters, and reduces to around 0.97 for various model
estimates, which is still far beyond the correct value.
It is known that the replacement of the overall $M_Q^5$ factor in
front of nonleptonic decay widths by the corresponding hadron mass $M_H^5$
provides a possible solution to the puzzle \cite{ac}, since
$(M_B/M_{\Lambda_b})^5=0.73$, $M_B$ and $M_{\Lambda_b}$ being the $B$ meson
mass and the $\Lambda_b$ baryon mass, respectively, is close to the
observed ratio. As pointed out in \cite{ae}, the same replacement also
explains the absolute $B$ meson decay rate, while the OPE approach using
the expansion parameter $M_b$ accounts for only 80\% of the decay rate.
However, the kinematic replacement for nonleptonic decays is at most an
{\it ad hoc} phenomenological ansatz without solid theoretical base.

In Ref.~\cite{ab} we have performed the PQCD
factorization of inclusive heavy hadron decays according to hadronic
kinematics, such that the above replacement is implemented naturally.
Generally speaking, a decay rate is expressed as the convolution of a hard
$b$ quark decay subamplitude with a universal heavy hadron distribution
function and several jet functions for light energetic final states. A jet
function contains double logarithms from the overlap of collinear and soft
enhancements in radiative corrections. For the semileptonic decays
$b\to cl{\bar \nu}$, because the final-state $c$ quark is massive, there
are no double logarithms and thus no jet functions. The transverse momentum
$p_\perp$ carried by the $b$ quark is introduced as a factorization scale,
above which PQCD is reliable, and radiative corrections are absorbed into
the hard subamplitude, and below which physics is regarded as being purely
nonperturbative, and absorbed into the distribution function.
To facilitate the factorization, we work in the impact
parameter $b$ space, which is the Fourier conjugate variable of $p_{\perp}$.

For the nonleptonic modes $b\to c{\bar u}d$ and $b\to c{\bar c}s$, which
involve more complicated strong interactions, additional subprocesses
need to be included except for those
in the semileptonic cases. These modes involve three scales: the $W$ boson
mass $M_W$, at which the matching condition of the effective Hamiltonian
for nonleptonic decays to the full Hamiltonian is defined, the
characteristic scale $t$ of the hard subamplitude, which reflects the
specific dynamics of heavy hadron decays, and the $b$ quark transverse
momentum $p_\perp$, or equivalently $1/b$, which serves as the
factorization scale mentioned above. To analyze the nonleptonic decays, we
have proposed a three-scale factorization formalism \cite{ah} that embodies
both effective field theory and the standard PQCD factorization theorems.
The hard gluon exchanges among quarks, which generate logarithms of $M_W$,
are factorized into a {\em harder} function \cite{ah}. Collinear and soft
divergences exist in radiative corrections simultaneously, when
the light final-state quarks, such as ${\bar u}$, $d$, and $s$,
become energetic. The resultant double logarithms $\ln^2({\bar
p}b)$, ${\bar p}$ being the jet momentum defined later, are
absorbed into the corresponding jet functions.

The various logarithmic corrections are summed to all orders using
the resummation technique and renormalization-group (RG) equations, whose
results are the evolutions among the scales $M_W$, $t$, and $1/b$.
The summation of the logarithms $\ln(M_W/t)$ is identified as the Wilson
coefficient of the effective Hamiltonian, which describes the evolution
from the characteristic scale $M_W$ of the harder function to the
characteristic scale $t$ of the hard subamplitude. The summation of the
logarithms $\ln(tb)$ leads to the evolution from $t$ to $1/b$. The double
logarithms $\ln^2({\bar p}b)$ in the jet functions are resummed to give a
Sudakov factor \cite{ab}, which suppresses the long-distance
contributions from the large $b$ region, and thus improves the
applicability of PQCD.

With the above ingredients, we have calculated the branching ratios of
the modes $b\to cl{\bar \nu}$, $b\to c{\bar u}d$ and $b\to c{\bar c}s$
for the inclusive $B$ meson decays \cite{ai}. It was observed that
the decay rates of the $b\rightarrow c\bar ud$ and $b\rightarrow c\bar cs$
modes are both enhanced by the QCD evolution effects.
In this way the semileptonic
branching ratio $B_{\rm SL}\equiv B(B\to Xl\bar\nu)$ is suppressed
without increasing the average charm yield $\langle n_c \rangle$ per $B$
decay. The controversy of the small $B_{\rm SL}$ in inclusive $B$ meson
decays, which cannot be explained by the quark model, is then resolved. This
solution differs from the usual attempts in the literature, in which the
$b\to c{\bar c}s$ mode is increased.
In this letter we shall further show that our formalism provides a
possible explanation  of the low lifetime ratio $\tau(\Lambda_b)/\tau(B_d)$.
Our predictions for the absolute decay widths of the $B$ meson and of the
$\Lambda_b$ baryon are also consistent with the data.

We quote the factorization formulas for the semileptonic and
nonleptonic $B$ meson decays \cite{ai}. The various momenta involved
in the semileptonic decays
$B(P_{B}) \rightarrow X_{c}+l(p_l)+\bar{\nu}(p_{\nu})$
are expressed, in terms of light-cone coordinates, as
\begin{equation}
 P_B=\left(\frac{M_B}{\sqrt{2}},\frac{M_B}{\sqrt{2}},0_{\perp}\right),
 \mbox{\ \ }
     p_l=(p^+_l,p_l^-,0_{\perp}),\mbox{\ \ }
     p_{\nu}=(p_{\nu}^+,p_{\nu}^-,{\bf p}_{\nu \perp}),
\end{equation}
where the minus component $p_l^-$ vanishes for massless leptons.
For convenience, we adopt the scaling variables,
\begin{equation}
x=\frac{2E_l}{M_B}\;,\mbox{\ \ } y=\frac{q^2}{M_B^2}\;,\mbox{\ \ }
y_0=\frac{2q_0}{M_B}\;,
\label{dl}
\end{equation}
with the kinematic ranges,
\begin{eqnarray}
2\sqrt{\alpha}\leq&x& \leq 1+\alpha-\beta,
\nonumber\\
\alpha\leq &y& \leq \alpha+(1+\alpha-\beta-x)
\frac{x+\sqrt{x^2-4\alpha}}{2-x-\sqrt{x^2-4\alpha}},
\nonumber\\
x+\frac{2(y-\alpha)}{x+\sqrt{x^2-4\alpha}}\leq &y_0&\leq 1+y-\beta,
\end{eqnarray}
where $E_l$ is the lepton energy and $q \equiv p_l+p_{\nu}$ the momentum of
the lepton pair. The constants $\alpha$ and $\beta$ are
$\alpha \equiv M^2_l/M^2_B$ and $\beta \equiv M^2_D/M^2_B$,
$M_l$ and $M_D$ being the lepton mass and the $D$ meson mass, respectively.
$M_D$ appears as the minimal invariant mass of the decay product $X_c$.

The $b$ quark momentum is parametrized as $P_b=P_B-p$, where
$p=((1-z)P_B^+,0,{\bf p}_\bot)$ is the momentum of the light cloud in the
$B$ meson. The factorization formula for the total inclusive semileptonic
decay width in the $b$ space is given by \cite{ai}
\begin{equation}
\frac{\Gamma_{\rm SL}}{\Gamma_0}=\frac{M_B^2}{2\pi}\int dxdydy_0
\int^1_{z_{\rm min}}{dz}
\int_0^\infty db b f_B(z){\tilde J}_c(x,y,y_0,z,b)H(x,y,y_0,z)\;,
\label{asb}
\end{equation}
with $\Gamma_0 \equiv (G_F^2/16\pi^3)|V_{cb}|^2M^5_B$. The momentum
fraction $z$ approaches 1 as the $b$ quark carries all the $B$ meson
momentum in the plus direction. The minimum of $z$, determined by the
condition $p_c^2 > M_c^2$, $p_c$ and $M_c$ being the $c$ quark momentum
and mass, respectively, is
\begin{equation}
z_{\rm min} =
\frac{\displaystyle\frac{y_0}{2}-y+\frac{M_c^2}{M_B^2}-
\frac{x}{\sqrt{x^2-4\alpha}} \left[-\frac{y_0}{2}+\frac{y}{x}+
\frac{\alpha}{x}\right] }
{\displaystyle 1-\frac{y_0}{2}-\frac{x}{\sqrt{x^2-4\alpha}}
\left[-\frac{y_0}{2}+\frac{y}{x}+\frac{\alpha}{x} \right] }.
\end{equation}

The function ${\tilde J}_c$ denotes the final-state cut on the $c$ quark
line, which is in fact part of the hard subamplitude.
$J_c$ and the lowet-order $H$ in momentum space are expressed
as \cite{ai}
\begin{eqnarray}
J_c&=& \delta\left( M_B^2\left\{z-(1+z)\frac{y_0}{2}+y+
\frac{x(1-z)}{\sqrt{x^2-4\alpha}}
\left[-\frac{y_0}{2}+\frac{y}{x}+\frac{\alpha}{x}\right]
\right \}\right.
\nonumber\\
& &\left. -M_c^2-{\bf p}^2_{\perp}\right),
\nonumber \\
& &\label{jc}\\
H&=&\left((y_0-x)\left\{1-\frac{(1-z)}{2}
\left(1-\frac{x}{\sqrt{x^2-4\alpha}}\right)\right\}
-\frac{(1-z)}{\sqrt{x^2-4\alpha}}
(y-\alpha)\right)\nonumber\\
&&\times \left(\frac{x}{2}\left\{1+z+(1-z)
\frac{\sqrt{x^2-4\alpha}}{x}\right\}
-y-\alpha\right).
\label{hi}
\end{eqnarray}
The universal $B$ meson distribution function $f_B$, determined from the
$B \rightarrow X_s \gamma$ decay, is given by \cite{ag},
\begin{equation}
f_B(z)= \frac{0.02647 z(1-z)^2}{[( z - 0.95 )^2 + 0.0034 z]^2}\;,
\label{bdf}
\end{equation}
which minimizes the model dependence of our predictions. We have ignored
the intrinsic $b$ dependence of $f_B$ from the infrared renormalon
contribution, since it does not affect our predictions of the decay
branching ratios. Note that the distribution function collects all-order 
soft gluons, which, being long-distance, are insensitive to the kinematics 
of the quarks they attach, such as masses. Hence, $f_B$ for the 
$b\to s\gamma$ and for $b\to c$ decays are the same. Through the $z$ 
dependence of $f_B$, the soft spectator effect is taken into account in our 
formalism.

Ignoring the penguin operators, the effective Hamiltonian
for the nonleptonic decay $b\to c{\bar c}s$ is
\begin{equation}
H_{\rm eff} = \frac{4G_{F}}{\sqrt{2}} V_{cb} V^{\ast}_{cs}
          [ \, c_{1}(\mu) O_{1}(\mu) +
            c_{2}(\mu) O_{2}(\mu) \, ]\;,
\label{eff}
\end{equation}
with $O_{1}=(\bar{s}_{L}\gamma_{\mu} b_{L})(\bar{c}_{L}\gamma^{\mu} c_{L})$
and $O_{2}=(\bar{c}_{L}\gamma_{\mu} b_{L})(\bar{s}_{L}\gamma^{\mu} c_{L})$.
For the $b\to c{\bar u}d$ mode, $V_{cs}$ and the ${\bar c}$ and $s$ quark
fields are replaced by $V_{ud}$ and the ${\bar u}$ and $d$ quark fileds,
respectively. It is simpler to work with the operators
$O_{\pm}=\frac{1}{2} (O_{2} \pm O_{1})$ and their
corresponding coefficients $c_{\pm} = c_{2}(\mu) \pm c_{1}(\mu)$, since
they are multiplicatively renormalized. The expressions of the
Wilson coefficients $c_\pm$
are referred to \cite{aj}.

To simplify the analysis, we route the transverse momentum $p_\perp$ of the
$b$ quark through the outgoing $c$ quark as in the semileptonic cases, such
that the $c$ quark momentum is the same as before. For kinematics, we make
the correspondence with the ${\bar c}$ (${\bar u}$) quark carrying the
momentum of the massive (light) lepton $\tau$ ($e$ and $\mu$) and with the
$s$ and $d$ quarks carrying the momentum of ${\bar\nu}$. The scaling
variables are then defined exactly by Eq.~(\ref{dl}). The factorization
formulas for the nonleptonic decay rates are quoted as \cite{ai}
\begin{eqnarray}
\frac{\Gamma_{\rm NL}}{\Gamma_0}
&=&\frac{M_B^2}{2\pi}\int dxdydy_0\int_{z_{\rm min}}^1 dz
\int_0^\infty bdb \left[ \frac{N_c\!+\!1}{2}
c_{+}^2(t) + \frac{N_c\!-\!1}{2} c_{-}^2(t) \right]
\nonumber \\
& & \times f_B(z){\tilde J}_c(x,y,y_0,z,b)H(x,y,y_0,z)
\nonumber\\
& &\times \exp\left[-\sum_j 2s({\bar p}_j,b)-s_{JS}(t,b)\right]\;.
\label{non}
\end{eqnarray}
The $B$ meson distribution function $f_B$, being
universal, is the same as that employed for the semileptonic decays.

The double logarithms contained in a light quark jet have been resummed
into the Sudakov factor $\exp \lbrack -2 s({\bar p}_j,b)\rbrack$
\cite{ab}, $j=s$ for the $b \rightarrow c\bar{c}s$ mode and
$j={\bar u}$, $d$ for the $b \rightarrow c\bar{u}d$ mode,
where the upper bound ${\bar p}_j\equiv (p_j^++p_j^-)$
of the Sudakov evolution is chosen as the sum of the longitudinal components
of $p_j$. We refer to \cite{WYL} for the expression of the exponent
$s$. In the numerical analysis below, $\exp(-2s)$ is set to unity as
${\bar p}_j< 1/b$. In this region the outgoing quarks are regarded as being
highly off-shell, and no double logarithms are associated with them. These
quark lines should be absorbed into the hard subamplitude, instead of into
the jet functions. The Sudakov factor, though important only in the
end-point region which is not essential for the calculation of total
decay rates, improves the applicability of PQCD as argued below.

Another exponential factor comes from the RG summation of
the single logarithms $\ln(tb)$, which is written as \cite{ab}
\begin{equation}
s_{JS}(t,b)=\int^{t}_{1/b}\frac{d\bar\mu}{\bar\mu}\left[
\sum_j\gamma_j(\alpha_s(\bar\mu))+\gamma_S(\alpha_s(\bar\mu))
\right]\;,
\end{equation}
$\gamma_j=-2\alpha_s/\pi$ and $\gamma_S=-C_F\alpha_s/\pi$ being the
anomalous dimensions of the light-quark jets and of the distribution
function, respectively. The summation of the single logarithms $\ln(M_W/t)$
leads to the Wilson evolution denoted by $c_\pm(t)$.
The scale $t$ is chosen as the maximal relevant scales,
\begin{equation}
t = \max \left[ {\bar p}_j, \frac{1}{b} \right]\;.
\end{equation}
Note that $t$, depending on the quark kinematics, probes the specific
dynamics of heavy hadron decays. That is, PQCD effects differ between the
$b\to c{\bar u}d$ and $b\to c{\bar c}s$ decays and between the $B$ meson
and $\Lambda_b$ baryon decays. The above single-logarithm evolutions play
an important role in enhancing the nonleptonic branching ratios and thus in
lowering the semileptonic branching ratios. This observation
is basically consistent with that in \cite{bagan}.
Since the large $b$ region is Sudakov
suppressed by $\exp(-2s)$ as stated before, $t$ remains as a hard scale,
and the perturbative evaluation of the hard subamplitude $H$ is reliable.

Without the PQCD evolutions, our formalism is equivalent to the HQET-based
OPE. This is obvious by expanding the distribution function $f_B$ in terms
of its moments \cite{ab} in the factorization formulas for the
semileptonic decays. $f_B$, though constructed on nonlocal oparators, is
constrained up to the second moment, the same as in the HQET approach.
The large logarithmic radiative corrections have been summed to all orders
into the evolution factors as shown in Eq.~(\ref{non}). The logarithms
$\ln(M_b/M_c)$ in the semileptonic decays are less important and neglected
as in Eq.~(\ref{asb}). The hard subamplitudes $H$ are the initial
conditions of the single-logarithm evolutions. In the HQET approach
radiative corrections are evaluated exactly to finite orders. Hence, the
difference between the two approaches resides only in the treatment of
PQCD effects. We shall show that PQCD effects, being process-dependent,
are responsible for the explanation of the lifetime ratio
$\tau(\Lambda_b)/\tau(B_d)$.

The above formalism can be generalized to the $\Lambda_b$ baryon decays
straightforwardly, for which the hard subamplitudes $H$, the final-state
cut $J_c$, and the Wilson coefficients $c_\pm$ remain the same but with the
kinematic variables of the $\Lambda_b$ baryon inserted. For the $\Lambda_b$
baryon distribution function, we assume the similar parametrization
\begin{equation}
f_{\Lambda_b}(z)=N\frac{z(1-z)^2}{[(z-a)^2+{\epsilon z}]^2}\;.
\label{eq: au}
\end{equation}
The moment associated with $f_{\Lambda_b}(z)$ need to be treated as a free
parameter, since spectra of $\Lambda_b$ baryon decays are still absent,
and $f_{\Lambda_b}$ has not been determined yet. The relations of the
parameters $N$, $a$ and $\epsilon$ to the first three moments of
$f_{\Lambda_b}(z)$ are demanded by the HQET-based OPE:
\begin{eqnarray}
\int_0^1 f_{\Lambda_b}(z) dz=1\;,
\;\;\;\;\int_0^1 (1-z)f_{\Lambda_b}(z) dz=\frac{{\bar\Lambda}(\Lambda_b)}
{M_{\Lambda_b}}+{\cal O}(\Lambda^2_{\rm QCD}/M_{\Lambda_b}^2)\;,
\nonumber \\
\int_0^1 (1-z)^2f_{\Lambda_b}(z) dz=\frac{1}{M^2_{\Lambda_b}}
\left({\bar\Lambda}^2(\Lambda_b)-\frac{1}{3}\lambda_1(\Lambda_b)\right)
+{\cal O}(\Lambda^3_{\rm QCD}/{M_{\Lambda_b}^3}),
\label{eq: ap}
\end{eqnarray}
with $\lambda_1(\Lambda_b)=
\langle \Lambda_b|{\bar b}(iD_\perp)^2b|\Lambda_b\rangle$ and
\begin{eqnarray}
M_{\Lambda_b}=M_b+{\bar\Lambda}(\Lambda_b)-
\frac{\lambda_1(\Lambda_b)}{2M_b}\;.
\label{eq: ar}
\end{eqnarray}
Note that $\lambda_1$ and $\bar\Lambda$ are not physically well defined
quantities due to the presence of infrared renormalon ambiguity. They may
bear very different values in different schemes.

We argue that only $\lambda_1(\Lambda_b)$ is a free parameter. The moments
of the $B$ meson distribution function obey a mass relation similar to
Eq.~(\ref{eq: ar}),
\begin{eqnarray}
M_B=M_b+{\bar\Lambda}(B)-\frac{\lambda_1(B)}{2M_b}\;.
\label{eq: arr}
\end{eqnarray}
Since the parameters ${\bar\Lambda}(B)=0.65$ GeV and $\lambda_1(B)=-0.71$
GeV$^2$ have been extracted from the available $B\to X_s \gamma$ spectrum
(see Eq.~(\ref{bdf})) and $M_B=5.279$ GeV is known, $M_b=4.551$ GeV is fixed
by Eq.~(\ref{eq: arr}). Substituting this $M_b$ into Eq.~(\ref{eq: ar})
with $M_{\Lambda_b}=5.621$ GeV, ${\bar\Lambda}(\Lambda_b)$ will be derived,
if $\lambda_1(\Lambda_b)$ is chosen. Using the values of
${\bar\Lambda}(\Lambda_b)$ and $\lambda_1(\Lambda_b)$, combined with the
normalization of $f_{\Lambda_b}(z)$, we obtain the corresponding
parameters $N$, $a$ and $\epsilon$.

Several reasonable choices of $\lambda_1(\Lambda_b)$ are made, and results
for $M_c=1.6$ GeV, $M_D=1.869$ GeV and $M_\tau=1.7771$ GeV are obtained.
Our predictions are not sensitive to the change of
$M_c$. The difference of the results for $M_c=1.5$ GeV from those for
$M_c=1.6$ GeV is less than 5\%. We adopt the leading-order Wilson
coefficients with the number of active quarks $n_f=5$. 
On the experimental side, the CLEO group reports
$B_{\rm SL}=(10.19\pm 0.37)\%$ and $\langle n_c\rangle=(1.12\pm 0.05)$
\cite{ak}, while the CERN $e^+e^-$ collider LEP measurements give
$B_{\rm SL}=(11.12\pm0.20)\%$ and $\langle n_c\rangle=(1.20\pm 0.07)$
\cite{LEP}. The lifetimes of the $B_d$ meson and of the $\Lambda_b$ baryon
are $\tau(B_d)=(1.56\pm0.06)\times 10^{-12}$ s and
$\tau(\Lambda_b)=(1.23\pm 0.05)\times 10^{-12}$ s, respectively,
from the LEP measurements \cite{ad}. Recent CDF result yields
$(1.32\pm0.16)\times10^{-12}$ s for the $\Lambda_b$ baryon lifetime
\cite{al}. The lifetime ratios $\tau(\Lambda_b)$/$\tau(B_d)$
are then $0.79\pm 0.06$ from LEP and $0.85\pm 0.11$ from CDF.

The conventional values of $\lambda_1$ obtained from QCD sum rules
are $\lambda_1^{\rm meson}\sim\lambda_1^{\rm baryon}=(-0.4\pm0.2)$ GeV$^2$.
Choosing $\lambda_1(\Lambda_b)=-0.4$ GeV$^2$ (with the correspondiong
${\bar\Lambda}(\Lambda_b)=1.03$ GeV), we derive the lifetime ratio about
0.81, which is much smaller than that from the OPE approach. If setting
$\lambda_1(\Lambda_b)$ to $\lambda_1(B)=-0.71$ GeV$^2$ (with the
correspondiong ${\bar\Lambda}(\Lambda_b)=0.99$ GeV) that was extracted from
the $B\to X_s \gamma$ decay, we arrive at a ratio 0.78 almost the same as
the central value of the LEP result. If further increasing the magnitude of
$\lambda_1(\Lambda_b)$ to $-1.1$ GeV$^2$ (with the correspondiong
${\bar\Lambda}(\Lambda_b)=0.95$ GeV), an even lower ratio about 0.75 is
reached. From the above analyses in a wide range of the HQET parameters,
it is obvious that the lower lifetime ratio is mainly due to the PQCD
effects. Our predictions for the absolute lifetimes of the $B$ meson and
of the $\Lambda_b$ baryon prefer a larger $|V_{cb}|$. Adopting
$|V_{cb}|=0.044$, we derive
$\tau(B_d)=1.56\times 10^{-12}$ s and
$\tau(\Lambda_b)=1.26\times 10^{-12}$ s, $1.22\times 10^{-12}$ s, and
$1.18\times 10^{-12}$ s for $\lambda_1(\Lambda_b)=-0.4$ GeV$^2$,
$-0.71$ GeV$^2$, and $-1.1$ GeV$^2$, respectively, which are
consistent with the data. We have adopted $|V_{cs}|=|V_{ud}|=1.0$ in
the present work.

In Table I we compare the branching ratio of each mode in the
$B$ meson and $\Lambda_b$ baryon decays in terms of the quantities
$\tau_{\tau\nu}=BR(b\to c\tau{\bar \nu})/BR(b\to cl{\bar \nu})$,
$\tau_{ud}=BR(b\to c{\bar u}d)/BR(b\to cl{\bar \nu})$, and
$\tau_{cs}=BR(b\to c{\bar c}s)/BR(b\to cl{\bar \nu})$, and observe
that $B_{\rm SL}$, $\langle n_c\rangle$ and all $\tau$'s except
$\tau_{ud}$ are enhanced a bit. It is interesting to examine these
tendencies in future experiments. The results listed in Table I are
insensitive to the variation of the distribution functions (less than 10\%
under 40\% variation of the moments), confirming that the branching ratios
are mainly determined by the PQCD effects. We emphasize that the predictions
for all the branching ratios of the $B$ meson decays are consistent with the
experiment data \cite{Ex}.

\vskip 0.5cm

We thank Prof. H.Y. Cheng for useful discussions. This work
was supported by the National Science Council of R.O.C. under
Grant Nos. NSC 88-2112-M-006-013 and NSC 88-2112-M-001-009.

\newpage

Table I. Comparision of $\tau_{\tau\nu}$, $\tau_{ud}$, $\tau_{cs}$,
$B_{\rm SL}$, and $\langle n_c\rangle$ in the $B$ meson and $\Lambda_b$
baryon decays for various $\lambda_1(\Lambda_b)$. The lifetime ratios
$\tau(\Lambda_b)/\tau(B_d)$ are also listed.

\vskip 0.5cm

\begin{center}
\begin{tabular}{ccccccc}
\hline
             & $\tau _{\tau \nu }$ & $\tau _{ud}$ & $\tau _{cs}$ &
$B_{\rm SL}$ & $\langle n_{c}\rangle $ & $\tau(\Lambda_b)/\tau(B_d)$ \\
\hline
$\lambda_{1}(B)=-0.71$ GeV$^2$ & 0.224 & 5.98 & 1.64 & 10.16 & 1.17 & \\ 
\hline\hline
$\lambda_{1}(\Lambda _{b})=-0.40$ GeV$^2$ & 0.254 & 5.14 & 1.75 & 10.94 &
1.19 & 0.81\\
\hline
$\lambda_{1}(\Lambda _{b})=-0.71$ GeV$^2$ & 0.261 & 5.26 & 1.68 & 10.87 &
1.18 & 0.78\\
\hline
$\lambda_{1}(\Lambda _{b})=-1.10$ GeV$^2$ & 0.266 & 5.36 & 1.68 & 10.76 &
1.18 & 0.75\\
\hline
\end{tabular}
\end{center}

\end{document}